\documentclass[english,tightenlines,a4paper,twocolumn]{revtex4}

\usepackage[pdftex]{graphicx}

\usepackage{natbib}
\usepackage{amssymb}
\usepackage{babel}

\graphicspath{{./Pics/}}

\begin{document}

\title{Defect-free ZnSe nanowire and nano-needle nanostructures}

\author{Thomas Aichele}
\email[Corresponding author. E-mail: ]{thomas.aichele@physik.hu-berlin.de}
\altaffiliation{present address: Institut f{\"u}r Physik, Humboldt Universit{\"a}t zu
Berlin, Hausvogteiplatz 5--7, 10117 Berlin, Germany}
\author{Adrien Tribu}
\author{Catherine Bougerol}
\author{Kuntheak Kheng}
\author{R\'egis Andr\'e}
\author{Serge Tatarenko}
\affiliation{Nanophysics and Semiconductor Group, CEA/CNRS/Universit{\'e} Joseph
Fourier, Institut N{\'e}el, 25 rue des Martyrs, 38042 Grenoble cedex 9, France}%

\date{\today}

\begin{abstract}
We report on the growth of ZnSe nanowires and nano-needles using molecular beam epitaxy (MBE).
Different growth regimes were found, depending on growth temperature and the Zn--Se flux ratio. By
employing a combined MBE growth of nanowires and nano-needles without any post-processing of the
sample, we achieved an efficient suppression of stacking fault defects. This is confirmed by
transmission electron microscopy and by photoluminescence studies.
\end{abstract}


\maketitle

\thispagestyle{headings}

Semiconductor nanowires (NWs) have attracted much attention in recent years because of
their properties and potential use in a variety of technological applications. NW
heterostructures are interesting candidates for the development of well-located and
size-controlled quantum dots (QDs) \cite{NW-QD}. Due to the narrow lateral size, QD
heterostructures in NWs can be directly grown on very defined positions and without the
necessity of self-assembly. This is an especially interesting feature for II-VI
materials, where self-assembled island formation occurs only within narrow windows of
growth conditions \cite{RobinAPL}. Recently, II-VI compound semiconductor NWs have been
synthesized by Au-catalysed metal-organic chemical vapour deposition (MOCVD) and
molecular-beam epitaxy (MBE) methods \cite{NW-MOCVD,NW-MBE}. An obstacle towards the
growth of optically active NW heterostructures are donor--acceptor pairs that form in
defects \cite{Philipose2}. These cause a strong spectral background which competes with
the excitonic emission in the QDs. Ref. \cite{Philipose1} reports the efficient reduction
of this spectral background after annealing of the NW samples in Zn-rich atmosphere.

When developing NW heterostructures, annealing or other post-growth processing of the
sample is often unfavourable as it may also influence the designed form of the
heterostructure through inter-diffusion of the constituents. Instead, growth methods,
that directly avoid the formation of defects are desired. In this letter, we report a
growth recipe for ZnSe NWs, where the amount of stacking fault defects is strongly
reduced. This was achieved by a combined growth of NWs and nano-needles without any
post-growth processing of the sample.

The ZnSe NWs were grown in the vapour-liquid-solid growth mode with gold particles as
catalysts. For comparison, GaAs(001) and Si(001) substrates were used. Degased surfaces
for MBE growth were obtained after annealing in ultra-high vacuum at 580$^\circ$ C. In
the case of GaAs, the effect of an epitaxial GaAs buffer layer was also investigated.
Interestingly the structural properties of the NWs depend very little on the utilized
substrate. Next, a thin gold film with thickness of 0.2--0.5 nm was deposited on the GaAs
substrates inside an electron beam metal deposition chamber. The gold film was dewetted
to a droplet-like surface by annealing the sample at $600^\circ$ C for 5 min. ZnSe MBE
growth was then performed with varying growth conditions. The sample transfers between
the MBE- and metal deposition chambers happened under ultra-high vacuum.

\begin{figure}[h]
\begin{center}\leavevmode
\includegraphics{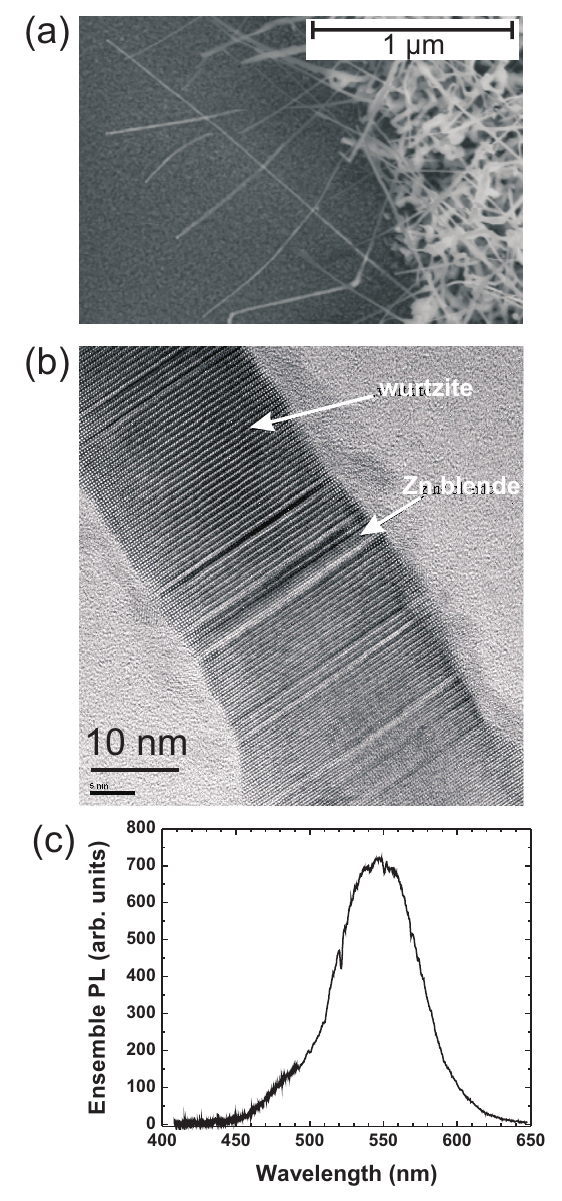}
\caption{Data obtained from NWs grown at 400$^\circ$ C: (a) SEM image of the as-grown sample at the
border of the gold-coated region. The right side shows the dense carpet of NWs. On the left,
individual NWs reach into the uncoated zone. (b) TEM image of a single NW. The arrows indicate
wurtzite and Zn-blende zones.}
\end{center}
\end{figure}

When growing under an excess of Se (Zn (Se) flux: 2.5 (7.5) $\times 10^7$~torr) and a
sample temperature of 350--450$^\circ$ C, a dense carpet of narrow NWs with high aspect
ratios covers the substrate. The NWs have a uniform diameter of 20--50~nm and a length up
to 2~$\mu$m after a growth of one hour (fig. 1(a)). Additionally to the NWs, the as-grown
substrate is covered with highly irregular nano-structures. Fig. 1(b) shows a
transmission electron microscopy (TEM) image of one NW. The crystal structure of the NWs
are predominantly wurtzite. However, the NWs are systematically intersected with regions
of zinc blende phase. We account for the formation of such defects and the presence of
highly irregular structures by non-ideal growth conditions at the initial stages of the
growth process. Possible reasons are the presence of non-uniform gold agglomerations
instead of small gold beads and the insertion of impurities during the gold deposition
process. The presence of both wurtzite and Zn-blende shows, that at the utilized growth
conditions both phases are allowed. Although, the observation of wurtzite structure is in
contrast to the Zn-blende that naturally occurs in bulk ZnSe, it is not an uncommon
behaviour for NWs, as discussed in ref. \cite{Glas_Wurtzite}.

\begin{figure}[h]
\begin{center}\leavevmode
\includegraphics{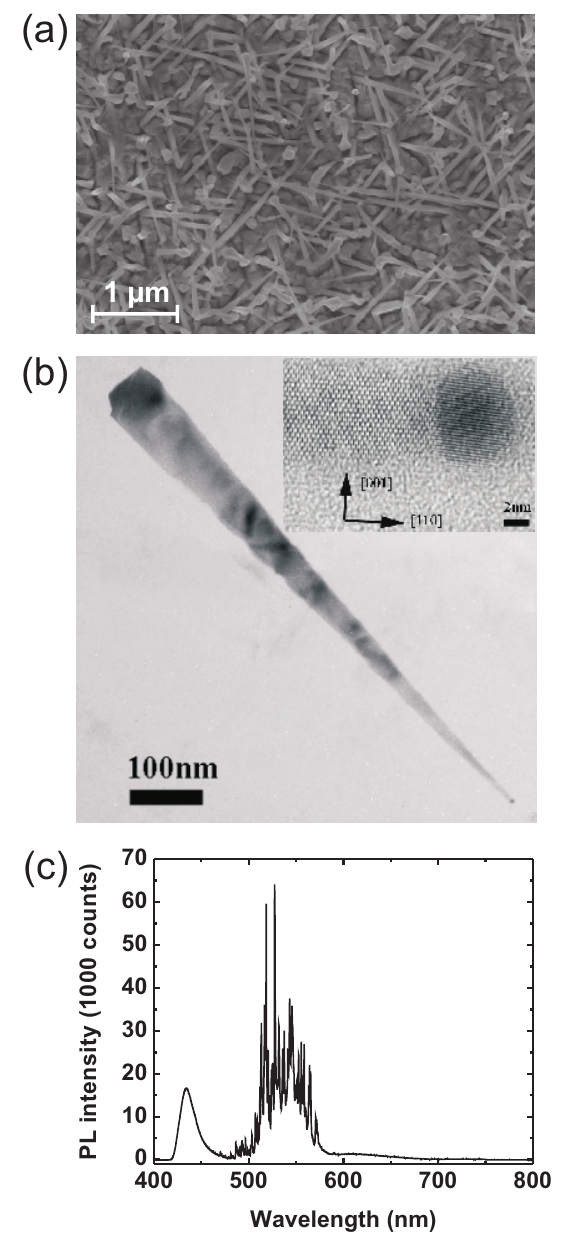}
\caption{Data obtained from nano-needles, grown at 300$^\circ$ C at Se excess: (a) SEM of the
as-grown sample. (b) TEM image of a single nano-needle. The inset shows a zoom to the region around
the tip.}
\end{center}
\end{figure}

When, on the other hand, growing at low temperature (300$^\circ$~C) or with inverted
Zn:Se flux ratio, needle-shaped NWs are formed (fig. 2(a)). Hereafter we will refer to
those nanostructures as {\em nano-needles} to distinguish them from the narrow {\em NWs}
described previously. By TEM (fig. 2(b)) we determined that the nano-needles have a wide
base (80 nm in diameter) and a sharp tip (5-–10 nm). We also observed darker and lighter
regions which again indicate the presence of stacking fault defects. The formation of
nano-needles instead of NWs is well accounted for by the slower adatoms mobility expected
at low temperature or at low Se flux. The slower mobility promotes nucleation on the
sidewalls before reaching the gold catalyst at the nano-needle tip. Moreover, we observed
that the nano-needles are predominantly wurtzite, as for NWs, but the wire axis is here
perpendicular to the c-axis instead of being parallel. In contrast also to the long NWs
(in fig. 1), the defect planes are here disoriented with respect to the nano-needle axis.
It seems that this disorientation hinders the propagation of defects in the growth
direction, especially for lower diameters. Defects zones are rapidly blocked on the side
walls, providing a high structural quality towards the nano-needle tip.

To carry out single-NW studies, the sample is put in a methanol ultra-sonic bath for 30~s in order
to detach some NWs from the substrate. Droplets of this solution are next placed on a fresh
substrate, leaving behind a low density of individual NWs. Ensemble spectra were taken on the
as-grown sample. All spectra in this paper were measured at a sample temperature of 5~K. The
photoluminescence (PL) of individual NWs were excited with a cw laser at 405~nm via a microscope
objective.

\begin{figure}[h]
\begin{center}\leavevmode
\includegraphics{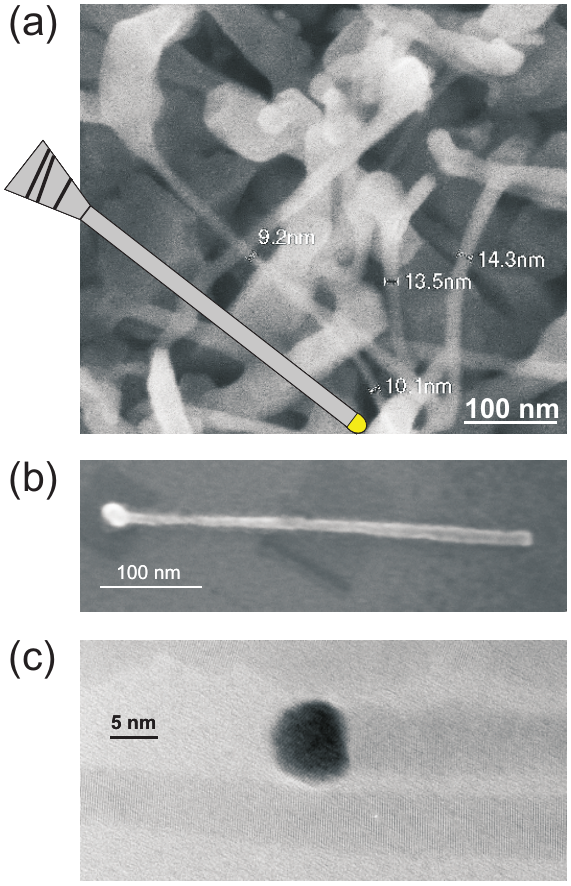}
\caption{Data obtained from combined growth of NWs on nano-needle tips. (a) SEM image of the
as-grown NW/nano-needle sample. (b) SEM of an isolated NW that broke off behind the thicker base.
(c) TEM image of two close-by NWs.}
\end{center}
\end{figure}

Both for NWs and nano-needles, we observe a broad spectral distribution within
500--600~nm, as seen in figs. 4(a) and (b). Even in the case of a single nano-needle, the
spectrum is dominated by many intense spectral lines, which we attribute to emission from
excitons localized at the defect zones in the NW \cite{Philipose2}. In contrast to the
observations in ref. \cite{Philipose1} on MOCVD-grown NWs, we do not see an enhancement
of the ZnSe band edge emission (443~nm at 5~K \cite{Malikova}) when growing under Zn-rich
conditions. The intense emission with 500--600~nm instead suggests that in both cases
(NWs and nano-needles), point defects effectively capture the excited charge carriers and
quench the band edge emission, as also reported in Ref. \cite{Xiang_noBE}. This is in
agreement with the high density of stacking fault defects observed by TEM.

\begin{figure}[h]
\begin{center}\leavevmode
\includegraphics{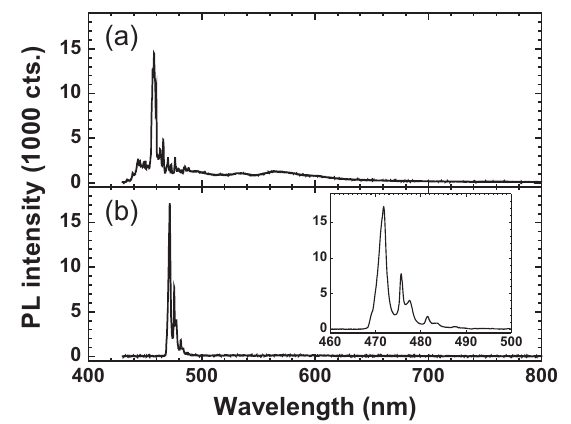}
\caption{PL spectra from the different samples: (a) NW sample from fig. 1; (b) nano-needle sample
from fig. 2; (c) and (d) combined NW/nano-needle sample from fig. 3. The inset of (d) is a zoom
into the region 460--500~nm.}
\end{center}
\end{figure}

The observation of a decreasing defect density from the base towards the top in the nano-needles
motivated us to modify the growth recipe in the following way: In the first partf, the sample is
grown with excess of Zn for 30~min, leading to the formation of nano-needle structures. Next, the
Zn- and Se-flux was inverted and NWs were grown for another 30 min on top of the nano-needles.
Thus, the growth at the side-walls was aborted and re-growth started on defect-free and
strain-relaxed nano-needle tips, where the high structural quality of the crystal lattice can be
preserved along the narrow NW that is now formed in this second growth step. Fig. 3 shows results
obtained from this sample. The structures have a broad base that tapers after a few ten nanometers
to thin NWs with thickness of 10--15~nm. As symbolized in the sketch in fig. 3(a) we expect that
stacking faults reduce towards the thin part of the NW, which is indeed the case, as seen in TEM
images of a single NW, fig. 3(c).

The suppression of defects has a strong effect on the PL of these nano-structures (fig
4). Due to the low density of  defects, the spectral emission between 500-600~nm that was
observed before from the samples in figs 4(a) and (b) is strongly reduced, leaving behind
only a small bunch of intense spectral lines between 450--500~nm. The weak and broad
distribution between 500--600 that remains in the ensemble PL, fig 4(c), is due to
excitons localized in defects in the thicker NW base. In the spectrum of a single NW that
broke off behind the thicker base, fig 4(d), this broad background is now globally
suppressed. In spite of this, no PL is observed at the ZnSe band-edge. A possible reason
is that, due to the very thin diameter of the NWs, additional surface states may form in
the bandgap \cite{Schmidt_Surface_InP_NWs} and introduce non-radiative decay channels
that quench the band-edge PL. The remaining narrow PL peaks around 470~nm in fig. 4(d)
can very likely be assigned to residual impurities responsible for donor-acceptor pair
emission and their related phonon replica \cite{Polimeni_ZnSeO}.

In summary we have  reported on MBE growth of ZnSe NWs. Depending on growth temperature
and Zn-Se flux ratio, we can tailor the structures between thin NWs with homogeneous
thickness of 20--50~nm, and nano-needles with broad base and a sharp tip with 5--10~nm
thickness. By combining nano-needle and NW growth, we achieved the growth of mostly
defect-free structures, without any post-growth treatment of the sample. This is
confirmed by TEM and PL measurements. In the latter, the emission of excitons localized
in the defect zones was strongly reduced. The suppression of defects is an important
pre-condition for developing QD heterostructures inside NWs. In a next step, we included
CdSe zones into the NW in order to form QDs. This will be reported in a subsequent
publication.

\section*{Acknowledgments}%

We are grateful to V. Zwiller for inspiring discussions. T.A. acknowledges support by
Deutscher Akademischer Austauschdienst (DAAD).

\newpage

\end{document}